\documentclass{svproc}
\usepackage{graphicx}%
\usepackage{multirow}%
\usepackage{amsmath,amssymb,amsfonts}%
\usepackage{mathrsfs}%
\usepackage[title]{appendix}%
\usepackage{xcolor}%
\usepackage{textcomp}%
\usepackage{manyfoot}%
\usepackage{booktabs}%
\usepackage{algorithm}%
\usepackage{siunitx}%
\usepackage{algorithmicx}%
\usepackage{algpseudocode}%
\usepackage{listings}%
\usepackage[symbol]{footmisc}

\usepackage[font=small,labelfont=bf]{caption}
\usepackage{subcaption}
\usepackage{url}

\usepackage{xurl}      % smarter URL line breaking

\usepackage{xcolor} % if not already loaded
\usepackage[
  colorlinks=true,
  linkcolor=blue,   % section/figure links (keep black if you want)
  citecolor=blue,    % <-- citations will be blue & clickable
  urlcolor=black      % URLs in blue
]{hyperref}

\newcommand{\swallow}[1]{}
\newcommand{\fullversion}[1]{}

\usepackage{todonotes}

\begin{document}
\mainmatter              % start of a contribution
\title{Embeddings of Nation-Level Social Networks}
\titlerunning{Embeddings of Nation-Level Social Networks}  
\author{%
  Tanzir Pial\inst{1}$^{\dagger}$%
  \and Flavio Hafner\inst{2,3}$^{\dagger}$%
  \and Dakota Handzlik\inst{1}%
  \and Enamul Hassan\inst{1}%
  \and Lucas Sage \inst{4}%
  \and Ana Macanovic \inst{5, 6}
  \and Tom Emery\inst{3}%
  \and Arnout van de Rijt\inst{5}%
  \and Steven Skiena\inst{1}%
}

\authorrunning{Pial et al.} % abbreviated author list (for running head)

\institute{
Stony Brook University, USA\\
\email{\{tpial,ehassan,skiena\}@cs.stonybrook.edu}
\and Netherlands eScience Center, The Netherlands
\and
Erasmus University Rotterdam, The Netherlands
\and
Toulouse School of Economics, Universite Toulouse Capitole, Toulouse, France 
% National Research Council of Italy, Italy
\and
European University Institute, Italy
\and
Utrecht University, The Netherlands
% \and
% Shahjalal University of Science and Technology, Sylhet, Bangladesh
}

\maketitle              % typeset the title of the contribution
% Place this right after \maketitle
\begingroup
  \renewcommand\thefootnote{\fnsymbol{footnote}}
  \footnotetext[2]{\ Equal contribution.}
% \endgroup
% \begingroup
  % \renewcommand\thefootnote{\fnsymbol{footnote}}
  % \footnotetext[1]{\ This research was conducted in whole or in part using ODISSEI, the Open Data Infrastructure for Social Science and Economic Innovations (https://ror.org/03m8v6t10) and was partially supported by NSF grants IIS1926781, IIS1927227, OAC-1919752, and a Fulbright Scholarship.}
  % \renewcommand\thefootnote{}% no automatic marker
  % \footnotetext{*This research was conducted in whole or in part using ODISSEI, the Open Data Infrastructure for Social Science and Economic Innovations (https://ror.org/03m8v6t10) and was partially supported by NSF grants IIS1926781, IIS1927227, OAC-1919752, and a Fulbright Scholarship.}
\endgroup
% \todo{Put in the full set of authors and connect with them.   I encourage you to be expansive, and if there are people you leave out you make sure they are OK with it.   Do this now.}
% put this in the preamble

% then right after \maketitle

\begin{abstract}
Full nation-scale social networks are now emerging from countries such as the Netherlands and Denmark, but these networks present challenging technical issues in working with large, multiplex, time-dependent networks.
We report on our experiences in producing dynamic node embeddings of the population network of the Netherlands.
We present (a) a layer-sensitive random walk strategy which improves on traditional flattening methods for multiplex networks, (b) a temporal alignment strategy that brings annual networks into the same embedding space, without leaking information to future years, and (c) the use of Fibonacci spirals and embedding whitening techniques for more balanced and effective partitioning.
We demonstrate the effectiveness of these techniques in building embedding-based models for 13 downstream tasks.
\swallow{
The abstract should summarize the contents of the paper
using at least 70 and at most 150 words. It will be set in 9-point
font size and be inset 1.0 cm from the right and left margins.
There will be two blank lines before and after the Abstract. \dots
% We would like to encourage you to list your keywords within
% the abstract section using the \keywords{...} command.
%\todo{You need a real abstract with keywords.}
}
\end{abstract}
\renewcommand\thefootnote{\arabic{footnote}} % (usually not needed if you used a group, but explicit is fine)
\setcounter{footnote}{0}                     % start body footnotes at 1

\section{Introduction}\label{sec1}

National statistical agencies and research institutions in countries like the Netherlands \cite{vdlaan2023}, Denmark \cite{cremers2025unveiling} and Sweden \cite{panayiotou2025anatomy} are beginning to assemble population-complete, nation-level social networks from administrative records (e.g., family links, co-residence, educational institution, workplace, and neighborhood). We work with the population network of the Netherlands, an annually-updated multiplex network constructed every year since 2009. As a representative example, the 2018 network contained roughly 17 million people (nodes) connected by 1.4 billion typed relationships (edges), for an average undirected degree of 82.3 edges per person \cite{vdlaan2023}.

Unlike platform-defined networks such as Facebook’s social graph \cite{ugander2011anatomy}, the administrative multiplex graph we work with links every person in the Netherlands through heterogeneous relation types, and evolves yearly as people are born, move, study, work, and form families. This unprecedented coverage enables longitudinal social analysis at a national scale. For example, Bokányi et al. (2023) \cite{bokanyi2023anatomy} examine degree, closure, and distance in this network, and García-Bernardo et al. \cite{garcia2024assessing} investigate SARS-CoV-2 transmission via school and family contacts using the same resource.

These national-level social networks are exciting resources for research, but working with them presents technical challenges on at least three different levels:

\begin{itemize}

% [Short version]
% \item 
% {\em Multiplex integration} -- The Netherlands network has five subnetworks (family, co-residence, education, work, neighborhood) with distinct tie semantics. Flattening these layers into a single set can blur these relationships and introduce bias, as seen in the DeepWalk \cite{perozzi2014deepwalk} embeddings recently trained by Lüken et al. \cite{lüken2025populationscalenetworkembeddingsexpose}.

\item
{\em Multiplex integration} -- The Netherlands population network contains five different subnetworks over the same vertex set/population, where each relation type (family, co-residence, education, workplace, neighborhood) encodes distinct tie semantics and degree patterns.
On this network, Lüken et al \cite{lüken2025populationscalenetworkembeddingsexpose} recently trained Deepwalk \cite{perozzi2014deepwalk} embeddings using flattening, which they used to predict voting behavior.
But flattening layers into a single edge set can blur these semantics, and create bias toward dense layers for random walk based computational methods.

A useful embedding should leverage layer information without overfitting to any one layer. 

% [Short version]
% \item {\em Temporal comparability} -- Embeddings from different years should be comparable, but independent training on separate networks results in non-overlapping spaces.

\item {\em Temporal comparability} -- It is desirable that vector representations of nodes representing people at different points in time be comparable, but embeddings trained independently on separate annual networks cannot be expected to reside in the same embedding space. 
Therefore, embedding similarity or distance metrics lose all meaning when comparing embedding spaces of temporally different networks.
Moreover, 
This requires the development of machine learning methods to unify embeddings trained separately from different years.

% [Short version]
% \item {\em Balanced node partitioning} -- Clustering the nation-scale network to achieve balanced partitions is challenging but necessary for machine learning tasks.

\item {\em Balanced node partitioning} -- Clustering nodes on a nation-scale network is challenging, particularly if we desire robust partitions into roughly equal-sized sets. Such partitions are useful when using cluster labels as features for machine-learning systems and in load balancing for parallel processing.

\end{itemize}

% [Done.]  \todo{AM: I wondered why the discussion of CBS access technicalities at this length came before the paper's contribution. Would pushing this part back make it more engaging for the readers who are now curious to hear what is it that we do?
% Tanzir: Fixed and moved.

% }

In this paper, we report on our experiences in creating effective yearly node embeddings of every Dutch citizen as part of a major project on predictability in the social sciences.
% while maintaining anonymity \todo{AM: Say whose anonymity and whose privacy concerns? Tanzir: removed.} and addressing privacy concerns. 
We adopt DeepWalk \cite{perozzi2014deepwalk} as a simple, scalable, task-agnostic backbone for general-purpose node embeddings, as opposed to heavier graph neural networks tuned to specific labeled tasks.  Along with other unsupervised alternatives, e.g. \cite{grover2016node2vec,tang2015line},
—node2vec (biased walks) \cite{grover2016node2vec}, LINE (first/second-order proximity) \cite{tang2015line}—
DeepWalk offers a strong trade-off between simplicity and scale.
Over the course of this project, we developed approaches to address each of the three challenges described above.
Our contributions here include:

\begin{itemize}
    \item {\em Layer-Aware Walks for Mulitplex Networks} -- As an alternative to flattening a $k$-layer multiplex network, we build layer-aware DeepWalk by interleaving layer tokens within biased random walks that stay in the current layer with a guaranteed probability of $p$, while distributing the remaining $(1-p)$ probability uniformly among all of the existing layers (including the current one). 
    We demonstrate that across all 13 downstream tasks that use these embeddings,
    DeepWalk with our layer-aware technique outperforms DeepWalk on the flattened network.
% \todo{A numerical claim here would be good -- on how many do we win?}

    \item {\em Temporal Embedding Alignment} -- To ensure longitudinal comparability, we project yearly embedding spaces into a common space (that of the earliest year) via a linear transformation. This alignment preserves the relational structure of the embeddings, which is confirmed by the high correlation 
    ($r > 0.5$) 
    between embedding cosine similarity scores from the original and aligned spaces. On this criterion, our method outperforms standard Orthogonal Procrustes alignment \cite{singer2019node}.
    % \todo{Give a concrete result here.}

    \item {\em Fibonacci Spirals for Node Partitioning} -- We demonstrate the effectiveness of SVD-based whitening to produce more isotropic embedding spaces as a preprocessing step before using Fibonacci spirals \cite{hannay2004fibonacci,swinbank2006fibonacci} to equi-partition the volume of the embedding space into easy-to-work-with cones.
    We show that this approach yields relatively balanced clusters that can integrate cleanly as features for machine-learning tools. 

\end{itemize}

The rest of this paper is structured as follows. Section \ref{sec:data} reports on the network dataset and downstream tasks that we employ in our studies.
Section \ref{sec:preparing-embeddings} discusses our algorithmic approaches in developing layer-aware node embeddings. We evaluate our embeddings on a set of 13 different social science tasks in Section \ref{sec:results}, to demonstrate the effectiveness of our techniques. Section \ref{sec:network-alignment} presents our rotation-based approaches for unifying the embeddings of a temporally dynamic network. Finally, Section \ref{sec:clustering} reports on graph partitioning methods using Fibonacci spirals and embedding whitening, before our conclusion.
% \todo{Make sure you have a reason for this figure -- if so explain, if not punt.}

% \begin{figure}
%     \centering
%     \includegraphics[width=0.5\linewidth]{fig/similarity-vs-hop-distribution.png}
%     \caption{Cosine-similarity distributions for node pairs grouped by shortest-path distance in the national multiplex network of 2020: 1-hop (green), 2-hop (blue), 3-hop (violet), and a random-pairs (red) baseline. Distributions shift left with increasing distance, confirming that the embeddings capture social proximity (1-hop highest similarity; random lowest).}
%     \label{fig:sim-vs-hop-dist}
% \end{figure}

%Here is \ref{fig:sim-vs-hop-dist}

\section{Data}
\label{sec:data}

For training the network embeddings, we use the population network of the Netherlands constructed from administrative registers by Statistics Netherlands (CBS) \cite{vdlaan2023}. For downstream prediction tasks we utilize the CBS registry data and the linked LISS survey data.
\subsection{A Whole Population Network}

The network data provides yearly snapshots starting from 2009. Each snapshot is a person–person multiplex graph over the de-identified resident population; nodes are individuals and edges are undirected relations typed by layer. The five relation layers present in the graph are:

% \todo{(maybe some more details on data privacy at CBS?)}

\begin{itemize}
    \item{\textbf{Family}}: Family ties are defined by blood or parental kinship, including parents, children, grandparents, grandchildren, and siblings (encompassing full, half, and unknown), as well as co-parents, aunts, uncles, nieces, nephews, and first cousins. Affinal ties, such as in-laws and step-relatives, are excluded.

\item{\textbf{Household}}: 
This layer links people living in the same household - i.e., at the exact same address. \fullversion{This also includes people living in the same institutional household (e.g., retirement home).}

% \todo{AM: I'd rephrase so that its clear: This layer captures people living in the same household - i.e., at the exact same address - with the focal person. This also includes people living in the same institutional household (e.g., retirement home)
% Tanzir: Done.}

\item{\textbf{Neighbors}}:
  All people living at the 10 geographically closest addresses and a maximum of 20 people living in an address within a 200-meter range are treated as neighbors. \fullversion{Additionally, a maximum of 20 people living in an address within a 200-meter range are also selected as neighbors. People living in institutions with more than 10 residents, such as nursing homes or prisons, are excluded. The layer is not symmetrical.}
  
  % \todo{AM: I don't think the second part is correct. For people living in apartments, their upstairs and downstairs neighbors are considered; further, a random sample of 20 people from the circle within 200 meters from their address it selected. Neighbors from institutional households (e.g., retirement homes) are not considered. Tanzir: Ana fixed it.}

\item{\textbf{Colleague}}:
This layer links employees who share an employer, with each person connected to at most the 100 geographically closest colleagues by residence to avoid extreme degrees in very large organizations.

\item{\textbf{Classmates}}:
Classmate ties are approximated through co-enrollment in an educational stratum, defined by a unique combination of school, location, track or level, and grade or year of study. \fullversion{This definition does not distinguish between multiple parallel classes within the same grade. The ties are permanent and are not deleted after graduation.} 
\end{itemize}

\begin{table}[t]%!h]
\centering
%\small
{\fontsize{7}{9}\selectfont
\begin{tabular}{@{} l l c l S[table-format=6.0] @{}}
\toprule
\multicolumn{1}{l}{\textbf{Code}} &
\multicolumn{1}{l}{\textbf{Prediction Task}} &
\multicolumn{1}{l}{\begin{tabular}[c]{@{}c@{}}R$^2$\\ (numeric)\\ \rule{8mm}{0.2pt}\\ AUC\\ (binary)\end{tabular}} &
\multicolumn{1}{l}{\textbf{Years}} &
\multicolumn{1}{l}{\begin{tabular}[c]{@{}c@{}}\textbf{Dataset}\\ \textbf{size}\end{tabular}} \\
\midrule

\multicolumn{5}{c}{\emph{Income (admin registers)}}\\
\cmidrule(lr){1-5}
INPBELI          & Taxable gross income                                                   & R$^2$ & 2022, 2023 & 191560 \\
INPPH780OUV      & \begin{tabular}[c]{@{}l@{}}Employer mandatory\\ pension contribution\end{tabular}    & R$^2$ & 2022, 2023 & 158170 \\
INPPG710PEN      & \begin{tabular}[c]{@{}l@{}}Employer supplemental\\ pension contribution\end{tabular} & R$^2$ & 2022, 2023 & 125180 \\
INPT5280PEN      & \begin{tabular}[c]{@{}l@{}}Retirement\\ pension income\end{tabular}                  & R$^2$ & 2022, 2023 & 12610  \\
SOCIAL\_SECURITY & \begin{tabular}[c]{@{}l@{}}Total social\\ security benefits\end{tabular}             & R$^2$ & 2022, 2023 & 152110 \\ 

\addlinespace[4pt]

\midrule
\multicolumn{5}{c}{\emph{LISS Survey (2020)}}\\
\cmidrule(lr){1-5}
% cv246            & VVD vote probability (\%)                                              & R$^2$ & 2020       & 1030  \\
% ch207            & \begin{tabular}[c]{@{}l@{}}Mental-health visits\\ in past year (\#)\end{tabular}     & R$^2$ & 2020       & 4990  \\
% cv248            & PVV vote probability (\%)                                              & R$^2$ & 2020       & 860   \\
cr166            & Self-identifies as Turkish                                             & AUC   & 2020       & 5450  \\
ca008            & \begin{tabular}[c]{@{}l@{}}Any car/boat/motor-cycle\\ ownership (last year)\end{tabular} & AUC   & 2020 & 5340  \\
cs039            & \begin{tabular}[c]{@{}l@{}}Donated to a religious\\ organization (last year)\end{tabular} & AUC & 2020 & 5530  \\
ch178            & \begin{tabular}[c]{@{}l@{}}On anxiety/depression\\ medication (current)\end{tabular} & AUC   & 2020       & 4990  \\
\addlinespace[4pt]

\midrule
\multicolumn{5}{c}{\emph{Demography (admin registers)}}\\
\cmidrule(lr){1-5}
Divorce          & \begin{tabular}[c]{@{}l@{}}Divorce/registered\\ partnership dissolution\end{tabular} & AUC   & 2021--2023 & 180950 \\
Union            & \begin{tabular}[c]{@{}l@{}}First marriage/registered\\ partnership (never-married)\end{tabular} & AUC & 2021--2023 & 200000 \\
Fertility        & Childbirth in year                                                                     & AUC   & 2021--2023 & 200000 \\
Twin             & Twin pair indicator                                                                     & AUC   &            & 36010  \\
\bottomrule
\end{tabular}
\caption{Thirteen downstream prediction tasks for evaluation}
\label{tab:pred-tasks}
}
\end{table}

\subsection{Downstream Tasks Data}

We evaluate the performance of our network embeddings in predictive tasks related to demographic events, income, and opinions (Table \ref{tab:pred-tasks}).

% We evaluate how well the network embeddings we generate perform in predictive tasks (summarized in Table \ref{tab:pred-tasks}) related to demographic events, income, and opinions, which are all important indicators widely used in social scientific research.

% \todo{AM: Could we start this section with a short intro so that readers get a smoother transition?; e.g.,: We evaluate how well the network embeddings we generate perform in predicting a variety of relevant individual characteristics and outcomes. This includes 13 tasks related to predicting demographic events, income, and opinions, which are all important indicators widely used in social scientific research.
% Tanzir: Done}

\begin{itemize}

\item{\em Income:} We predict five income-related variables for 2022 and 2023, excluding non-positive incomes and applying log-transformation.

% \item{\em Income:} We predict five income-related variables for the years 2022 and 2023. We exclude individuals with non-positive incomes and do a log-transformation on the variables. All reported results are on the log-transformed variables.

\item{\em LISS survey response:} We predict answers to four 2020 survey questions on mental health, identity, and behavior from the LISS panel \footnote{\url{https://www.lissdata.nl/} (accessed Dec 2024)}.

% \item{\em LISS survey response:} We predict answers to four survey questions from 2020 spanning topics such as mental health medication, identity, and behavior. We use the answers from The Longitudinal Internet Studies for the Social Sciences (LISS) panel\footnote{\url{https://www.lissdata.nl/} (accessed Dec 2024)}. \fullversion{a probability-based, nationally representative survey administered by Centerdata (Tilburg University) across the Netherlands since 2007 with monthly online questionnaires and periodic refreshment samples to maintain representativeness~\cite{scherpenzeel2011data}.}

% \todo{AM: I'd suggest rephrasing this, e.g.: We predict answers to 7 survey questions from 2020 spanning topics such as voting intention, health, identity, and behavior. We use the answers from The Longitudinal Internet Studies for the Social Sciences (LISS) panel(accessed Dec 2024), a probability-based, nationally representative survey administered by Centerdata (Tilburg University) across the Netherlands since 2007....
% Tanzir: Done.
% }

\item{\em Demographic events:} We predict divorce, union, and fertility events for 2021-2023. All datasets are built via random sampling to preserve population base rates.

% \item{\em Demographic events:} We predict several relevant demographic events. 
% For \textit{divorce}, we predict dissolution among couples who were in a marriage or a registered partnership on 31st December 2020. For \textit{union}, we predict marriages/registered partnerships for previously never married individuals (age $\ge 18$)  between 2021-2023. For \textit{fertility}, we predict whether an individual will have a child in the same time period. All demographic datasets are built via random sampling that preserves the population base rates. 

% \todo{AM: I find that the word "label" comes up all of a sudden here. Maybe we can say "We predict dissolution of marriage or a registered partnership after... For union, we predict marriages/registered partnerships for previously never married individuals between 2021-2023, and for fertility we predict whether an individual will have a child in the same time range.
% Tanzir: Done}

We consider the \textit{twin indicator} task as a high-signal control. For every positive pair (known twins), this task samples a negative pair of unrelated individuals at random. A simple distance-based classifier should succeed here because twins are directly connected through the \textit{Family} layer as siblings; expected strong performance thus can sanity-check that the embeddings preserve close-kin structure. 

% \todo{AM: the family layer is not very clear to me. Tanzir: Added as siblings.}

\end{itemize}

% \todo{These tasks and table should be lifted to an appendix}

\section{Layer-aware Node Embeddings}\label{sec:preparing-embeddings}

\fullversion{A random walk consists of a sequence of nodes in a graph where successive nodes are directly connected and the next node at each point is picked at random. DeepWalk interprets random walks across a graph as sequences of words and then applies the Word2Vec algorithm \cite{mikolov2013efficient} to produce node (word) embeddings. Word2Vec learns embeddings so that words occurring in similar contexts share similar vectors. In the graph domain, this translates naturally: nodes that frequently co-occur in random walks -- i.e., nodes that are closely connected -- will be embedded close together. We tackle a more complex problem of generating embeddings for a multiplex network here\footnote{The code is available here:  \url{https://github.com/odissei-lifecourse/layered_walk}}. 
For efficient data processing, we parallelize the algorithm in Python with Numba. To train the embeddings on the generated random walks, we use software from \cite{wang2019dgl}.
}

A random walk consists of a sequence of nodes in a graph where successive nodes are directly connected, and the next node at each point is picked at random. DeepWalk maintains that nodes that frequently co-occur in random walks -- i.e., nodes that are closely connected -- will be embedded close together. We tackle a more complex problem of generating embeddings for a multiplex network here.\footnote{Our code is available here:  \url{https://github.com/odissei-lifecourse/layered_walk}} 

\subsection{Layer Persistence}

In our multiplex network, each edge belongs to one of five layers. Performing a union of the edge-sets in all layers (i.e., flattening the layers into one network) is a simple strategy to utilize all the data at our disposal. This creates an undesired effect where successive nodes in a walk may be connected through different layers. As an example, the colleague of a person's sister will be equally likely to be part of the walk as the sister's husband. We posit that this will lead to less meaningful walks and co-occurrence in the walks will represent less tight-knit network relationships.

% \todo{AM: I am not sure I understand what "tight-knit co-occurrence" means exactly.
% Tanzir: Modified to make more sense}

To tackle this problem, we modify the probabilities of the next node in the walk; with probability $p$, we choose a neighbor from the same layer as the latest layer used, and with the remaining $(1-p)$ probability, we resample layers, choosing uniformly among all the layers in which the current node has an edge (which may sample the current layer again). With our intuitive choice of $p=0.8$, there is at least a 64\% chance that any contiguous four-node subsequence in a walk remains within the same layer, enabling the model to learn cohesive, layer-specific patterns while still benefiting from cross-layer transitions. We create DeepWalk embeddings combining this flattening of graph and layer persistence technique. The walks still lack explicit edge-type information, therefore we denote the embeddings as the layer-blind embeddings. We treat the layer-blind embeddings as a baseline.

% \todo{AM: It's a bit unclear to me what "balance" means here. Tanzir: Ignored}

% \todo{AM: Maybe it'd be helpful to explain which part related to "flattening" (I guess they 1-p part?) and which to "persistence" (the p part?) Tanzir: Ignored}.

\subsection{Interleaving Hub Nodes}
DeepWalk does not account for edge types, so we introduce “hub” nodes to preserve edge-type information, adopting the idea from \cite{dong2017metapath2vec}. For example, a $A \rightarrow B \rightarrow C$ becomes $A \rightarrow [\text{\textit{Family}}] \rightarrow B \rightarrow [\text{\textit{Education}}] \rightarrow C$, ensuring the embedding process captures relationship types ($A\leftrightarrow B$ -- family, $B \leftrightarrow C$ -- education). This approach is referred to as layer-aware embeddings.

 % DeepWalk, by design, has no notion of what kind of edge links two nodes. To preserve edge-type data, we adapt an idea from \cite{dong2017metapath2vec}. Instead of connecting two people directly, we insert a “hub” node representing the edge type between them. For example, a walk segment $A \rightarrow B \rightarrow C$, where $A\leftrightarrow B$ is a family layer edge and $B \leftrightarrow C$ is an education layer edge, becomes $A \rightarrow [\text{\textit{Family}}] \rightarrow B \rightarrow [\text{\textit{Education}}] \rightarrow C$. This ensures the embedding process incorporates the types of relationships. We refer to this as layer-aware embeddings. 
 
\begin{figure}[t]
    \centering
    \includegraphics[width=\linewidth]{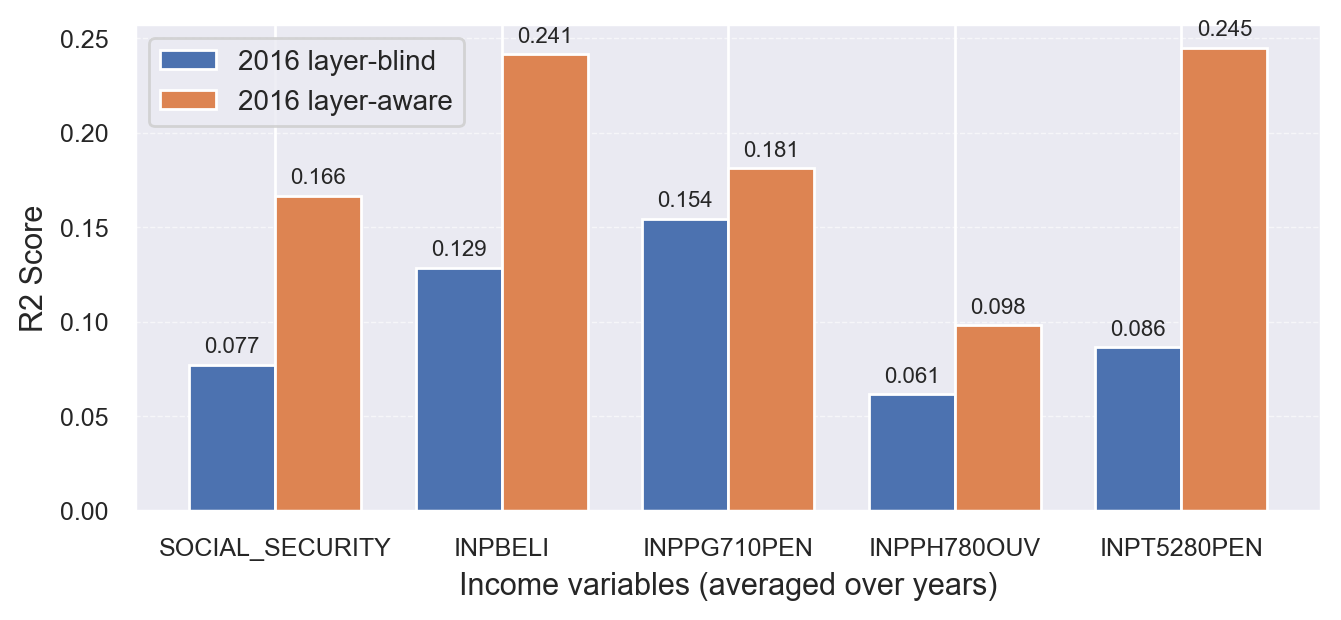}
    \caption{Layer-aware embeddings outperform the layer-blind embeddings for all 5 income variables. Here the performance is averaged over two years for each variable.}
    \label{fig:income-layer-aware-vs-blind}
\end{figure}

 Figures \ref{fig:income-layer-aware-vs-blind} and \ref{fig:AUC-binary} show that layer-aware embeddings outperform layer-blind embeddings across all 13 downstream tasks in Table \ref{tab:pred-tasks}. We used the following parameters for both embeddings: layer-persistence $p=0.8$, walk length $L=40$, walks per node $n=4$. Layer-blind walks consist of $L$ person-nodes, while layer-aware walks alternate between person nodes and hub nodes. Both embeddings were trained for 50 epochs with an output embedding dimension of 128.

\section{Downstream Predictive Tasks}\label{sec:results}

We evaluate the predictive utility of our graph embeddings on the 13 downstream tasks detailed in Section \ref{sec:data}. For each task, we partition the data into training (70\%), validation (10\%), and test (20\%) splits. All predictive models are implemented as a 2-layer feed forward neural network. 

\noindent\textbf{Embedding Sets:} Though we have created embeddings for every year, we use three sets of embeddings in these experiments: 2016 layer-blind, 2016 layer-aware, and 2020-layer-aware embeddings. We choose these years particularly to highlight how effective these embeddings are 4-7 years into the future, as well as to investigate how more recent embeddings perform compared to older embeddings.

\noindent\textbf{Input Representation:} For 11 of the 13 tasks, the model input is the graph embedding of a single individual. The two relational tasks—divorce prediction and twin pair identification—take as input the concatenated embeddings of the two focal individuals.

\noindent\textbf{Training Procedure:} For binary classification tasks, we minimize a weighted (to mitigate class imbalance) binary cross-entropy loss. For regression tasks, the loss function is Mean Squared Error. For all tasks, we perform a hyperparameter search over batch size $\in \{16, 128\}$ and learning rate $\in \{1e-3, 5e-4, 5e-5\}$. Each model is trained for a maximum of 20 epochs with early stopping (patience=2) based on validation performance. The final model for each task is selected based on optimal performance on the validation set.

\noindent\textbf{Evaluation:} Model performance is reported on the held-out test set. For binary classification tasks, we report the Area Under the ROC Curve (AUC). For regression tasks, we report the coefficient of determination ($R^2$). 

\fullversion{
\todo{AM: I'd probably order them as: INPBELI, INPPG710PEN, INPPH780OUV, SOCIAL\_SECURITY, INPT5280PEN, to go from more general -> employment related -> social security general -> pension}
}

\begin{table}[h]
\setlength{\tabcolsep}{10pt}
\centering
%{\fontsize{8}{10}\selectfont
\small
\begin{tabular}{l|rrr}
target                            & \multicolumn{1}{l}{year} & \multicolumn{1}{l}{\begin{tabular}[c]{@{}c@{}}2016\\ layer-aware\end{tabular}} & \multicolumn{1}{l}{\begin{tabular}[c]{@{}c@{}}2020\\ layer-aware\end{tabular}} \\ \hline
\multirow{2}{*}{SOCIAL\_SECURITY} & 2022                      & 0.175                                 & \textbf{0.206} \\
                                  & 2023                      & 0.158                                 & \textbf{0.186} \\ \hline
\multirow{2}{*}{INPBELI}          & 2022                      & \textbf{0.261}                        & 0.258          \\
                                  & 2023                      & \textbf{0.222}                        & 0.206          \\ \hline
\multirow{2}{*}{INPPG710PEN}      & 2022                      & \textbf{0.183}                        & 0.176          \\
                                  & 2023                      & \textbf{0.179}                        & 0.175          \\ \hline
\multirow{2}{*}{INPPH780OUV}      & 2022                      & 0.099                                 & \textbf{0.113} \\
                                  & 2023                      & 0.097                                 & \textbf{0.106} \\ \hline
\multirow{2}{*}{INPT5280PEN}      & 2022                      & \textbf{0.277}                        & 0.253          \\
                                  & 2023                      & 0.213                                 & \textbf{0.215}
\end{tabular}
\caption{The 2020 layer-aware embeddings perform similarly to 2016 layer-aware embeddings for predicting income variables in 2022 and 2023. Performances decrease from 2022 to 2023 for all embedding-target pairs. }
\label{tab:income-preds}
%}
\end{table}

\subsection{Predicting Future Income}

Our income prediction analysis reveals three key findings. First, Figure \ref{fig:income-layer-aware-vs-blind} shows that 2016 layer-aware embeddings outperform layer-blind embeddings across both prediction years and income variables, highlighting the value of encoding edge-type information.
Second, Table \ref{tab:income-preds} shows a decline in performance when predicting income for 2023 compared to 2022, likely due to increased uncertainty and unobserved variables.
Finally, we observe no clear performance advantage when using the more recent 2020 layer-aware embeddings over the 2016 layer-aware set (Table \ref{tab:income-preds}). Contrary to the intuition that a more contemporary graph should better reflect an individual's future financial trajectory, the differences in $R^2$ appear marginal.

This finding raises a compelling question: does the network structure over a 4-year period exhibit insufficient change to materially benefit later-year embeddings, or does our embedding methodology fail to effectively capture these temporal dynamics? A promising direction for future work is to stratify the population by the volatility of their local network structure. Comparing the predictive performance of 2016 vs. 2020 embeddings for individuals with highly stable networks versus those with highly dynamic networks as well as the actual income distribution of these two groups could reveal the coupling between network evolution, income prediction, and model capability.

% \begin{figure}[t]
% \begin{subfigure}{0.5\textwidth}
%     \centering
%     \includegraphics[width=\linewidth]{fig/figureA_overlay_auc_v2.png}
%     \label{fig:auc-scores}
% \end{subfigure}
% \begin{subfigure}{0.5\textwidth}
%     \centering
%     \includegraphics[width=\linewidth]{fig/figureA_overlay_r2_v2.png}
%     \label{}
% \end{subfigure}
% \caption{Caption}
% \end{figure} 

\begin{figure}[t]
\centering
\includegraphics[width=0.9\linewidth]{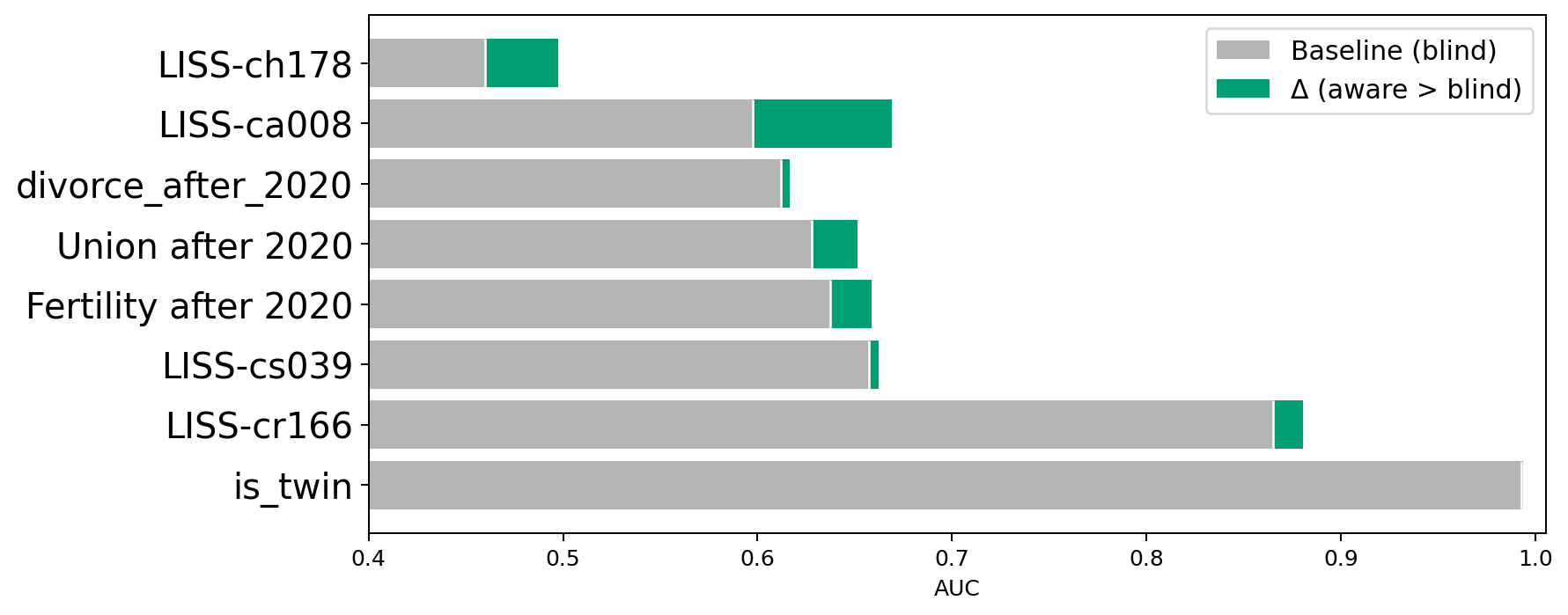}
\caption{AUC of embeddings vs different binary classification tasks. The grey bars represent AUC of the baseline layer-blind embeddings and the green portions represent the improvement achieved by layer-aware embeddings.}
\label{fig:AUC-binary}
\end{figure}

% \todo{I cannot understand this figure \ref{fig:r2-scores}.   Make these big enough to understand and explain -- maybe 1.0*textwidth each or one of them -- but you need to explain.}

% \todo{AM: I am a bit lost as to why we discuss the 2016 embeddings here and not e.g., 2020 ones? (bc of LISS?) The intro suggests that we'd build embeddings for all years since 2009, but it might be worth to say why we only present 2016 and 2020.  Tanzir: Because we still want to show that layer-aware embeddings are better than layer-blind embeddings. I have added a paragraph at the beginnning of section 6 discussing why we do these experiments with only 3 sets of embeddings}

% \todo{AM: Being rather nitpicky, I'd say this could go as a first task we present throughout the paper if it serves as a sanity check. But it was listed as the last one in the earlier sections on tasks. Tanzir: Ignoring for now}

\subsection{Performance on Survey and Demographic Tasks}

% \todo{AM: perhaps "classification tasks"? Tanzir: Fixed}

The predictive performance on all binary classification tasks is presented in Figure \ref{fig:AUC-binary}, revealing several key insights. The 2016 layer-aware embeddings yield improvement in AUC over the layer-blind baseline across all classification tasks. This reinforces the value of encoding edge semantics, as the model successfully leverages the distinct relational information from different network layers. 

% \todo{AM: Or perhaps  "suggesting the information relevant for predicting this medication use is not inherently present in or captured by the network structure" - otherwise, it sounds slightly redundant to say that anxiety med information is not present in the network structure. Tanzir: Done}

% \todo{AM: If you ask me, the high performance on the Turkish self-identification question is a much better example, even though the performance difference is relatively smaller. I think the car indicator is more likely mirroring the INPBELI difference; but we do not discuss the INPBELI performance difference in such detail. 
% Tanzir: Ignoring}

Task-specific performance meets expectations. We achieve near-perfect AUC (0.99) on the twin pair prediction task, validating our embeddings' ability to preserve familial relationships. Predictions for personal attributes like anxiety medication use (LISS-ch178) show near-random performance (AUC $\sim$0.5), indicating the network structure doesn’t capture relevant information. In contrast, layer-aware embeddings yield a 10\% AUC improvement for vehicle ownership prediction (LISS-ca008). For example, current colleagues are stronger indicators of asset ownership than former classmates, a distinction only the layer-aware model captures. The strong performance in predicting self-identification as Turkish (AUC $>$ 0.8, LISS-cr166) likely reflects tight-knit ethnic communities within family and neighborhood networks, captured by both embedding types.

\section{Temporal Alignment of Embeddings}\label{sec:network-alignment}

The next problem we tackle is the temporally dynamic nature of the graphs. When we generate a separate set of vertex embeddings for each year's network, these embeddings reside in different latent spaces because of their independent training processes. To enable a unified analysis across time, such as clustering individuals across all years, we must align these yearly embedding spaces into a single common frame of reference.

We perform this alignment by projecting the embeddings from each subsequent year ($t > 2009$) back into the embedding space of a base year, 2009. This choice prevents information leakage from future years into past representations. For each target year and for each dimension $i$ of the embedding space, we learn a separate linear regression model which uses the full embedding vector of an individual in the target year as the predictor to estimate their corresponding value in dimension $i$ of the individual's 2009 embedding.

% \todo{AM: What's the outcome being predicted and what are the features?
% Tanzir: Modified the last line to make more sense.}

% This model maps the full embedding vector of an individual in the target year to their corresponding value in dimension $i$ within the 2009 base space.

% Concretely, for each dimension $i$, the features are the individual’s entire embedding vector in year $t$ ($\mathbf{x}_t\in\mathbb{R}^d$), and the outcome is that same individual’s scalar coordinate in the 2009 space along dimension $i$ ($y^{2009}_i$); the fitted linear model $f_i(\mathbf{x}_t)\approx y^{2009}_i$ provides the mapping.

% \todo{AM: Sorry for my ignorance, but why is this important? Tanzir: Ignored. Explained in the next paragraph}

\begin{figure}[t]
\centering
\includegraphics[width=0.9\linewidth]{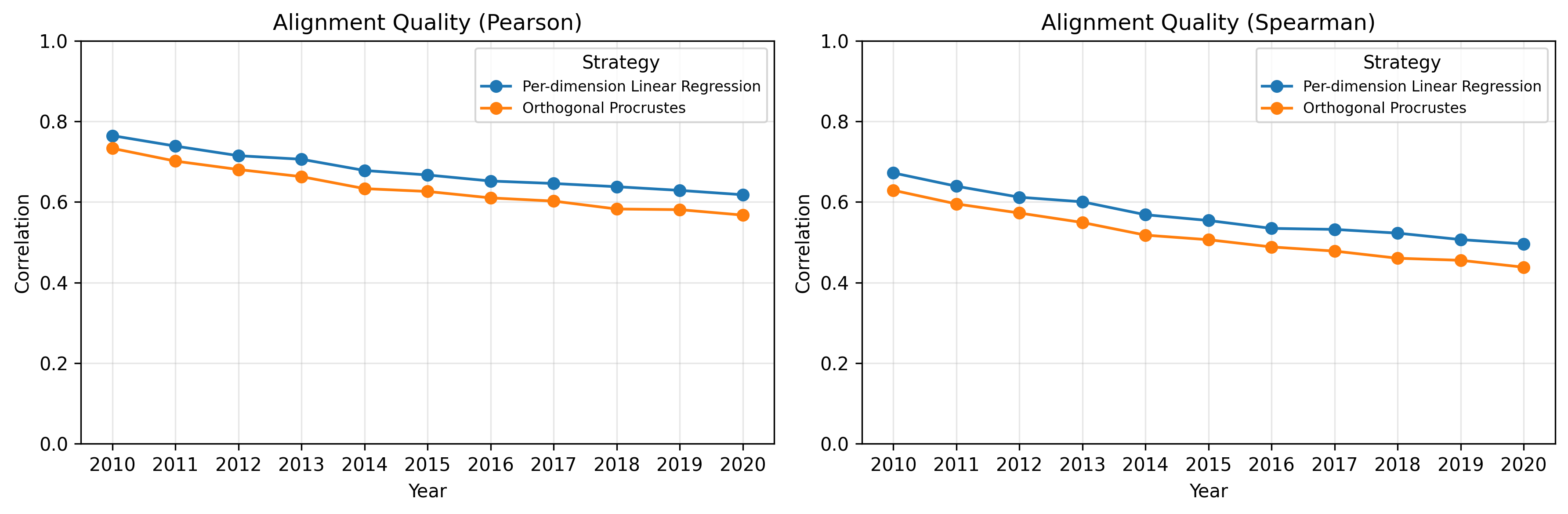}
\caption{Alignment quality over time when mapping embeddings from each source year (2010--2020) to the 2009 target space. We compare Per-dimension Linear Regression (OLS) and Orthogonal Procrustes. Panels show Pearson (left) and Spearman (right) correlations between aligned and target embeddings (higher is better). Performance declines with temporal distance; OLS consistently outperforms Orthogonal Procrustes.}
\label{fig:alignment-quality}
\end{figure}

We compare our method with Orthogonal Procrustes analysis \cite{singer2019node}, which applies a single orthogonal transformation per year to minimize distortion. While Procrustes preserves pairwise cosine similarities, our linear regression method offers more flexibility by learning a separate map for each dimension. We evaluate both methods on a validation set of 132,530 random pairs, computing the cosine similarity of their embeddings before and after alignment. As shown in Figure \ref{fig:alignment-quality}, the per-dimension linear regression consistently achieves higher correlation across all years, demonstrating its superior ability to preserve the relational structure. Thus, we select linear regression as our preferred method for temporal alignment.

% We compare our method with Orthogonal Procrustes analysis \cite{singer2019node}, which learns a single orthogonal transformation per year to minimize distortion. While Procrustes preserves pairwise cosine similarities, our linear regression method offers more flexibility by learning a separate map for each dimension. We evaluate both methods on a validation set of 132,530 random pairs, computing the cosine similarity of their embeddings before and after alignment. The correlation between pre- and post-alignment similarity scores is then measured.

% We contrast this approach with Orthogonal Procrustes analysis \cite{singer2019node}, a common alignment technique that learns a single orthogonal transformation (rotation/reflection) per year to minimize the overall distortion. \fullversion{While Procrustes preserves pairwise cosine similarities for the train set by design, our linear regression method offers greater flexibility by learning a distinct linear map for each dimension.}
% We compare the two methods on how well they preserve the embedding relationships for a validation set of 132,530 random pairs of individuals for each target year. For each pair, we computed the cosine similarity of their embeddings both before and after alignment using both methods. We then measured the correlation between these pre- and post-alignment similarity scores. 

\section{Equipartitioning Node Embeddings}\label{sec:clustering}
A primary application of network embeddings is their use as feature vectors in downstream predictive models. However, the high dimensionality of these continuous representations can complicate their direct integration. We posit that clustering serves as an effective method for feature reduction in this context, transforming each embedding into a discrete cluster identifier. While this is a lossy transformation, it facilitates compatibility with a wide array of machine learning tools that operate efficiently on categorical inputs. Furthermore, the resulting clusters often reveal patterns among groups within the data, providing valuable insight into the latent structure captured by the embeddings.

% \todo{AM: I am not sure what this means, but it might be because of my lack of familiarity with the field. Tanzir: Agreed. Made the wording simpler.} 

\fullversion{
\begin{figure}[ht]
    \centering
    \includegraphics[width=0.8\linewidth]{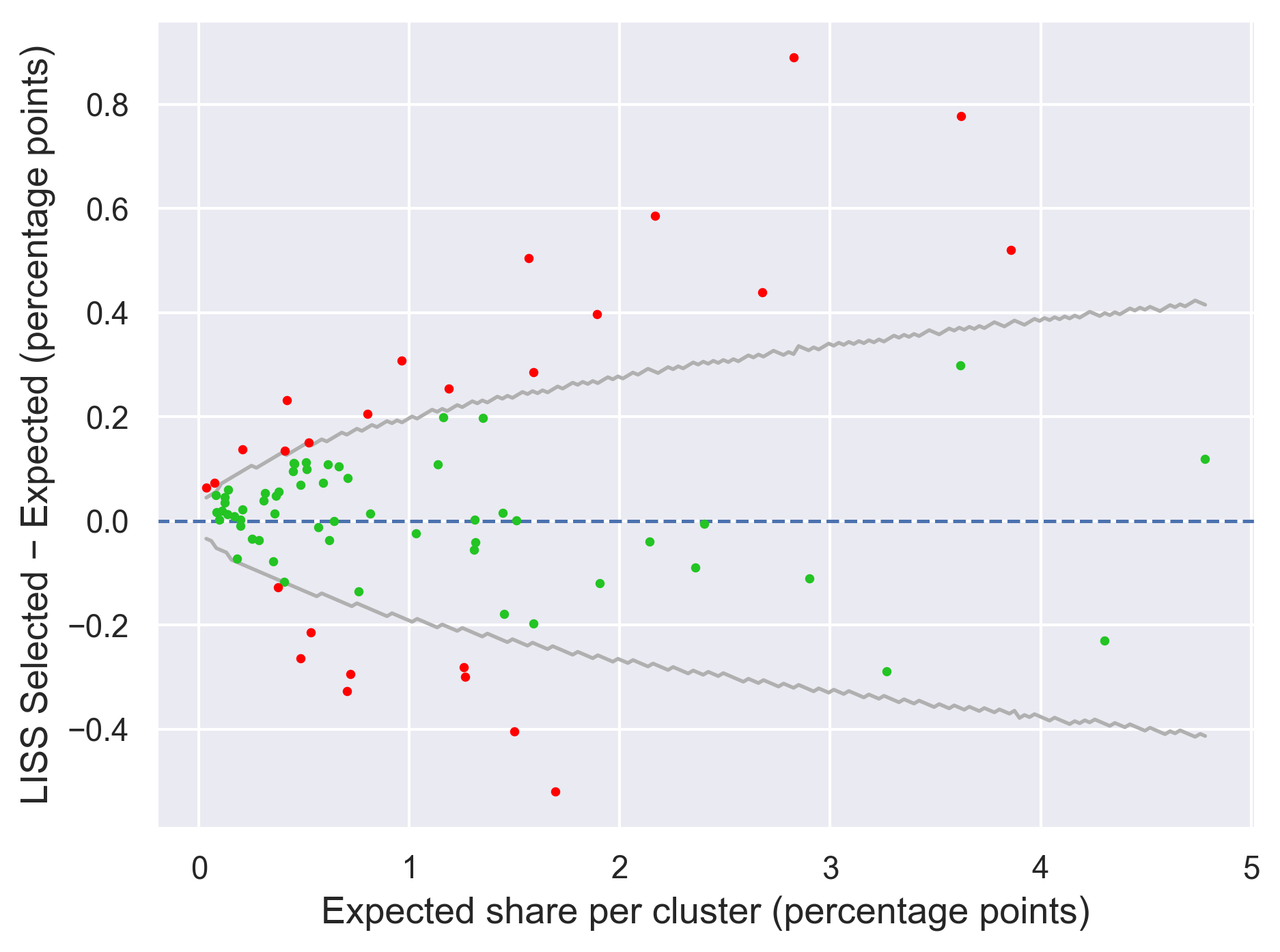}
    \caption{Difference-vs-expected plot for cluster representation in the LISS survey panel. The solid envelope depicts 95\% reference limits from a hypergeometric model (randomly drawing $x$ individuals from population of $n$ without replacement). Points inside (green) are compatible with chance; outside (red) indicate deviations beyond expected sampling noise ($p < 0.05$). 26 out of the 83 clusters show disproportionate representation in the LISS panel. 17 of the 100 clusters are removed as they had less than 10 LISS survey participants, which goes against CBS privacy rules.}
    \label{fig:LISS-Clusters}
\end{figure} 
}

\begin{figure}[t]
%\centering
\begin{subfigure}{0.38\textwidth}
    \raggedright
    \includegraphics[width=\linewidth]{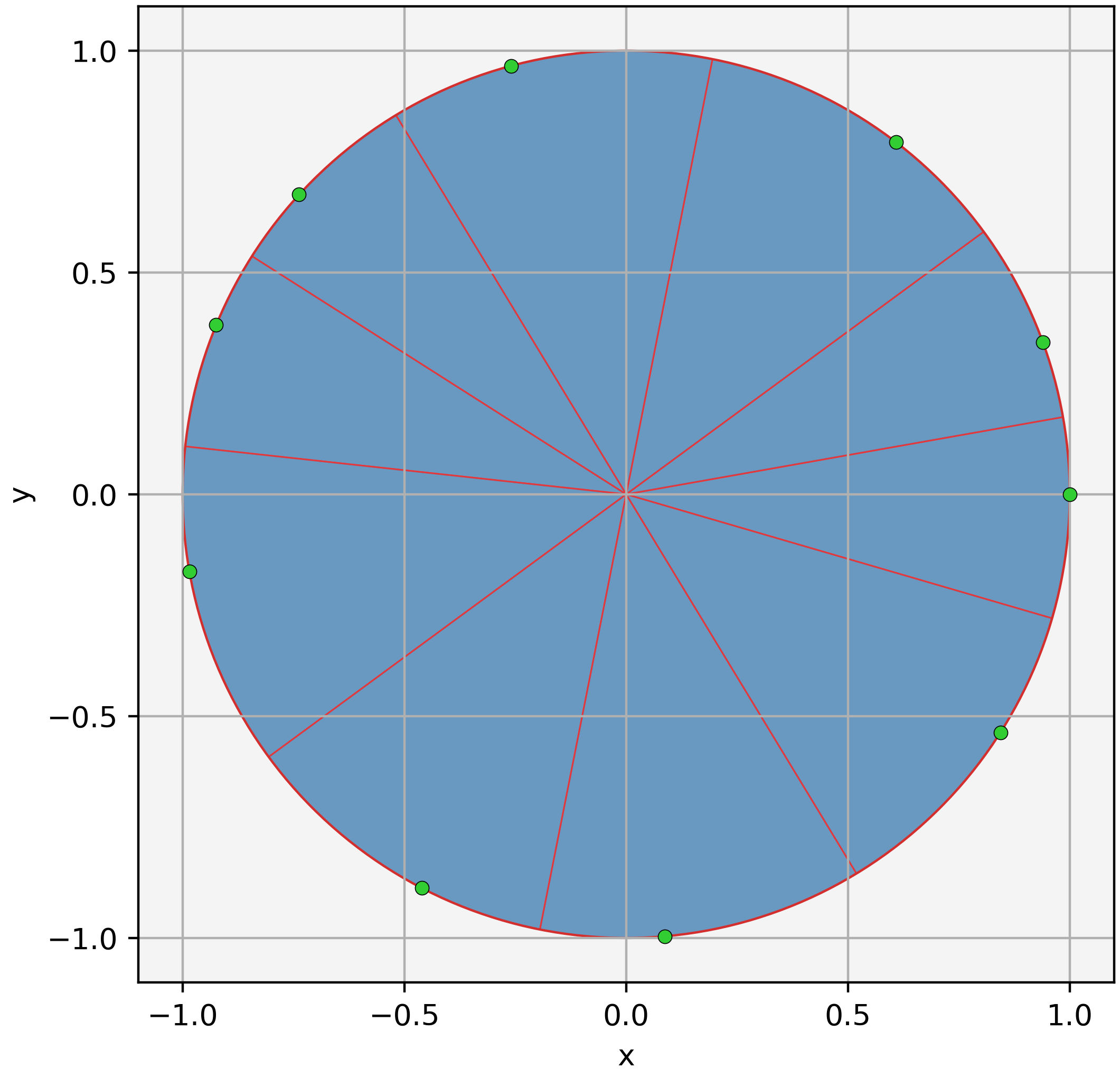}
\end{subfigure}\hfill
\begin{subfigure}{0.42\textwidth}
    \raggedleft
    \includegraphics[width=\linewidth]{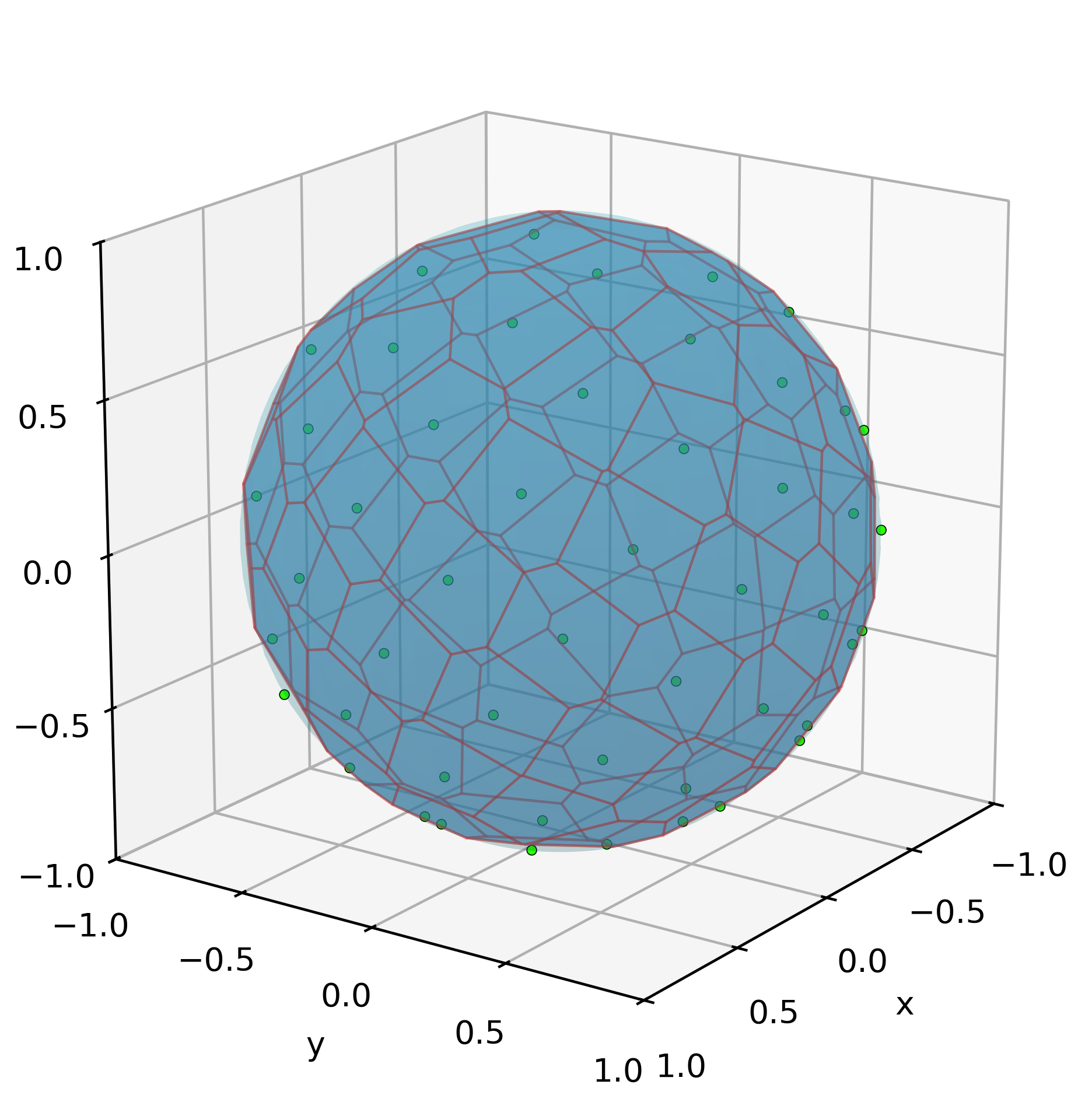}
\end{subfigure}
\caption{Example Fibonacci grids with Voronoi cells in 2D and 3D. On the left, the 2D cells are generated with 10 points on the Fibonacci lattice that divides the circumference into 10 almost equal parts. On the right, 100 points generate the 100 Voronoi clusters on the 3D sphere's surface. Fibonacci grids in higher dimension can generate similar Voronoi cells.}
\label{fig:fibo-grids}
\end{figure}

\subsection{Fibonacci Grids}

Traditional clustering algorithms, such as $k$-means, are data-dependent, meaning the discovered cluster structure is intrinsically tied to the input dataset. In contrast, we employ a fixed, data-independent clustering scheme that partitions the embedding space into $k$ pre-defined cells of (approximately) equal volume. This approach enables the direct comparison of different embedding methods by evaluating their distributions across an identical set of clusters, providing a measure of how uniformly they occupy the latent space.

We define clusters using $k$ direction vectors, assigning each embedding to the nearest vector by cosine similarity. This partitions the space into $k$ spherical Voronoi cells. The challenge then is to generate well-distributed directions on the surface of a $d$-dimensional sphere.

 % We define our clusters using $k$ direction vectors: any embedding is assigned to the cluster whose direction vector is its nearest neighbor, as measured by cosine similarity. This partitions the space into $k$ spherical Voronoi cells. Then the remaining challenge is to generate $k$ well-distributed directions or points on the surface of a $d$-dimensional unit sphere.

Generating $k$ evenly spaced points on a 3d sphere is a classical problem known as the Tammes problem. For values of $k \le 14$ an exact solution is known; for larger $k$, the Fibonacci lattice offers an excellent and computationally efficient approximation in 3d \cite{swinbank2006fibonacci,saff1997distributing,hannay2004fibonacci}. We leverage the generalized Fibonacci grid method for higher dimensions \cite{purser2008generalized} to generate our set of $k$ direction vectors in $\mathbb{R}^d$. Figure \ref{fig:fibo-grids} shows two example Fibonacci grids with the Voronoi cells in 2D and in 3D. 

f
\subsection{Whitening for Isotropic Embedding Spaces}\label{subsec:whitening}

Node embeddings are often anisotropic, causing an imbalanced distribution across clusters. This concentration of embeddings in a small subset of clusters diminishes their utility as discrete features, as it limits any model's ability to discriminate among large number of individuals in the same cluster. To promote isotropy, we apply embedding whitening, a post-processing technique that transforms the embeddings to have zero mean and unit covariance. This is achieved by centering the data and applying a linear transformation derived from PCA, resulting in uncorrelated dimensions of unit variance. Whitening is known to improve performance in embedding-based language models \cite{huang2021whiteningbert} for some tasks. 
%Fig. \ref{fig:whitening-cdf} shows that whitening effectively balances the distribution of embeddings across the 100 Fibonacci clusters we generated.

\begin{figure}[t]
\begin{subfigure}{0.45\textwidth}
\raggedleft
\includegraphics[width=\linewidth]{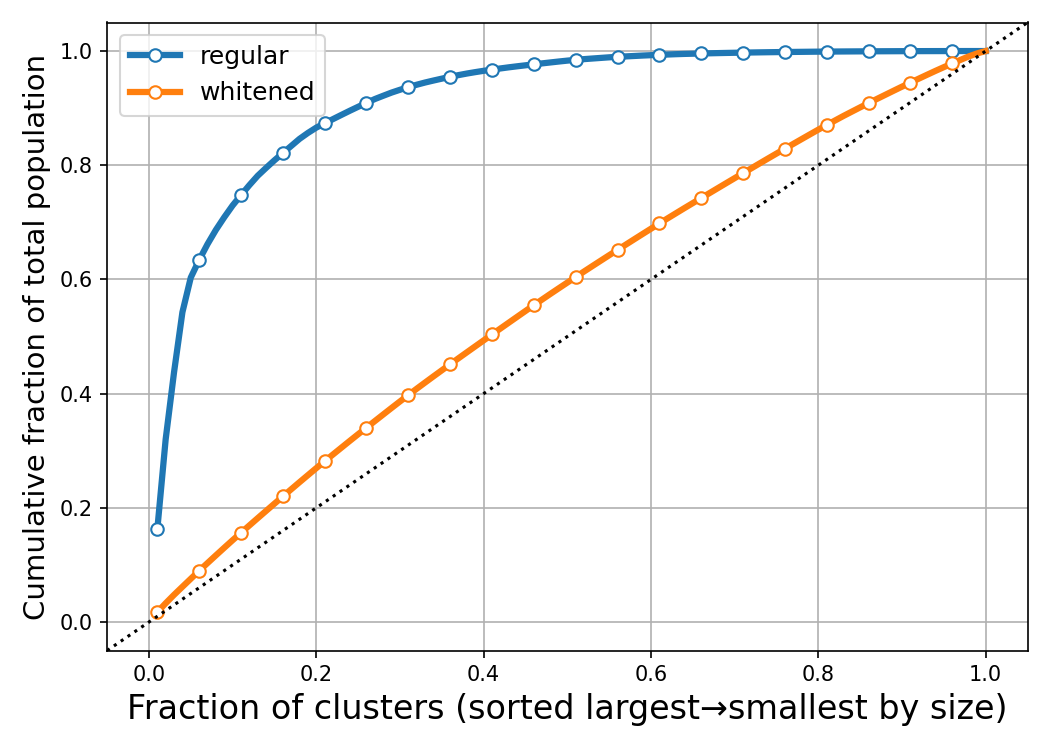}
\caption{}
\label{fig:whitening-cdf}
\end{subfigure}\hfill
\begin{subfigure}{0.55\textwidth}
\raggedright
\includegraphics[width=\linewidth]{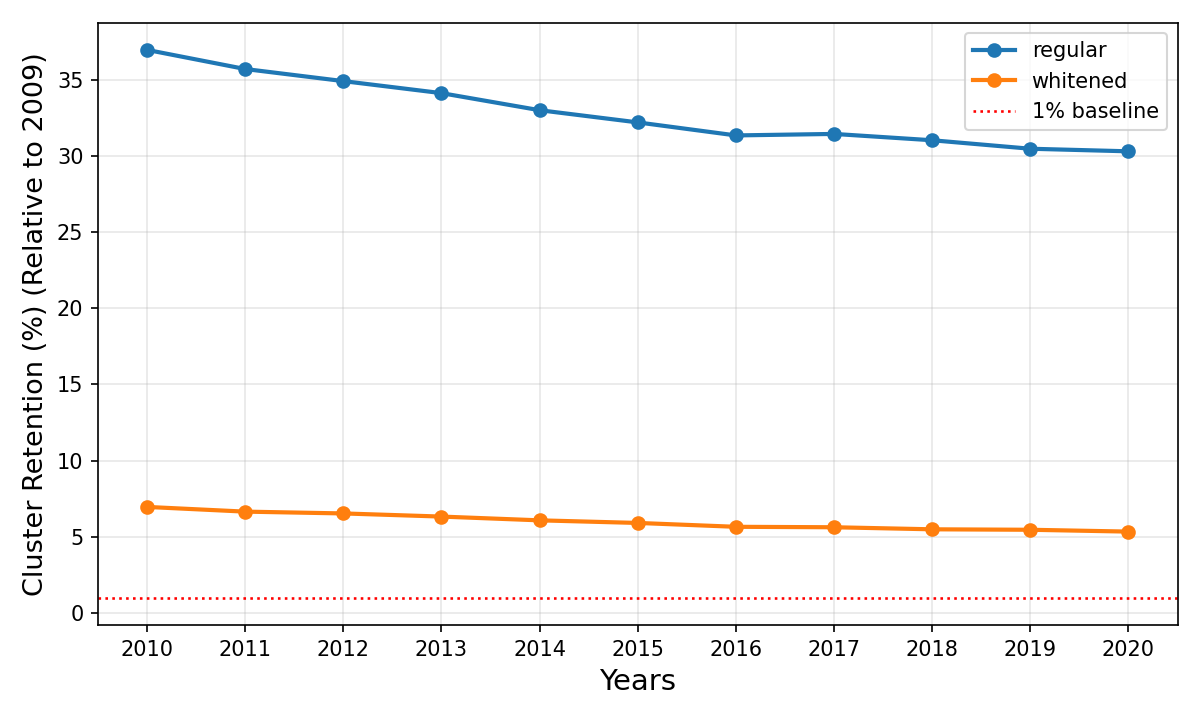}
\caption{}
\label{fig:cluster-retention}
\end{subfigure}
\caption{(a) CDF of cluster sizes (every fifth cluster marked). Whitening yields near-uniform distribution (near the $y=x$ line) compared to regular embeddings, where less than 5\% of the clusters contain more than 50\% of the regular embedding points. (b) Fraction of individuals remaining in their 2009 cluster.}
\end{figure}

A critical concern is that whitening may disrupt the relational structure of the embeddings. To investigate, we compare pairwise cosine similarities of 100,000 random person embeddings before and after whitening and temporal transformation. Pearson correlation for embeddings from 2010 to 2020 ranges from 0.44 to 0.52 (median = 0.48), while for 2009 (whitening only) it’s 0.70, indicating some distortion from each transformation. We also evaluate temporal stability by measuring, for the 13 million individuals in every annual graph (2009–2020), the fraction that remains in its 2009 cluster each year. High values would indicate embeddings fail to reflect temporal change, while values near 1\% would signal instability (random baseline around 1\%). As shown in Fig.~\ref{fig:cluster-retention}, temporally aligned (non-whitened) embeddings retain 30–35\%, while whitening reduces this to 5–7\%, still above the random baseline, indicating meaningful evolution over time, with a gradual decline suggesting stable tracking of societal dynamics.

% A critical concern is whitening may disrupt the relational structure of the set of embeddings. We investigate this by comparing pairwise cosine similarities of 100,000 random person embeddings before and after whitening and temporal transformation. For embeddings from 2010 to 2020, the Pearson correlation ranges from 0.44 to 0.52 (median = 0.48); for the base year 2009 (whitening only), it is 0.70, indicating that each transformation introduces some additional distortion.

% We then evaluate temporal stability by measuring, for the $13$ million individuals present in every annual graph in 2009–2020, the fraction that remains in its 2009 cluster in each subsequent year. Too high values would suggest embeddings that fail to reflect temporal change, while values near 1\% would indicate excessive instability (the random baseline is $\sim1\%$ with 100 equal-size clusters). As shown in Fig.~\ref{fig:cluster-retention}, temporally aligned (non-whitened) embeddings retain 30--35\% in the same cluster (anisotropy being one of the factors behind the high value), while whitening reduces this to 5--7\%—still well above the random baseline and indicative of meaningful year-to-year evolution. The decline with time is gradual, suggesting the embeddings track societal dynamics without becoming unstable.

% \vspace*{-0.5\baselineskip}        % tighten space *before* the 

\section{Conclusion}

% \vspace*{-0.495\baselineskip}        % tighten space *before* the 

We have demonstrated effective techniques for building versatile and high-quality node embeddings from dynamic, nation-scale multiplex social networks.
We have shown that (a) layer-aware embeddings of multiple networks outperform layer-blind embeddings over 13 predictive tasks (b) back rotation to the origin year effectively unifies embeddings from different years, and (c) Fibonacci spirals coupled with whitening techniques offer a straightforward way to partition network nodes into comparable-sized clusters. 
Future work will investigate how these embeddings can improve the performance of machine-learning models utilizing sociological covariates.  
% ... end of Conclusion
% \vspace*{-0.5\baselineskip}        % tighten space *before* the heading
{\small
\section*{Acknowledgements}
This research was conducted in part using ODISSEI, the Open Data Infrastructure for Social Science and Economic Innovations (https://ror.org/03m8v6t10), with analysis performed using Microdata from Statistics Netherlands (CBS) under project number 9424. This work was supported by NSF grants IIS1926781, IIS1927227, OAC-1919752, and a Fulbright grant. Funding from the French Agence Nationale de la Recherche (under the Investissement d’Avenir programme, ANR-17-EURE-0010) is gratefully acknowledged. We thank ARP Maranca (Princeton University) for the twin-pair indicator dataset.
}

% \vspace*{-0.9\baselineskip}        % tighten space after the paragraph , the Open Data Infrastructure for Social Science and Economic Innovations 
% The analysis was conducted using Microdata from Statistics Netherlands in project number 9424

% \section*{Acknowledgements}

\bibliographystyle{spmpsci} % We choose the "plain" reference style
\bibliography{refs} % Entries are in the refs.bib file
\end{document}